\documentclass[preprint,showpacs,preprintnumbers,amsmath,amssymb,natbib,nofootinbib]{revtex4}

\usepackage{graphicx}
\usepackage{epsfig}
\usepackage{dcolumn}
\usepackage{bm}
\usepackage[utf8]{inputenc}
\inputencoding{utf8}

\newcommand\trick[1]{}

\def\be{\begin{equation}}
\def\ee{\end{equation}}

\def\ba{\begin{eqnarray}}
\def\ea{\end{eqnarray}}

\begin{document}
\title{The Layzer-Irvine equation in theories with non-minimal coupling between matter and curvature}

\author{Orfeu Bertolami\footnote{E-mail: orfeu.bertolami@fc.up.pt} and Cláudio Gomes\footnote{E-mail: claudio.gomes@fc.up.pt}}
\affiliation{Departamento de F\'isica e Astronomia, Faculdade de Ci\^encias da Universidade do Porto, Rua do Campo Alegre s/n, 4169-007 Porto, Portugal}
\affiliation{Centro de Física do Porto, Rua do Campo Alegre s/n, 4169-007 Porto, Portugal}

\date{\today}

\begin{abstract}
We derive the Layzer-Irvine equation for alternative gravitational theories with non-minimal coupling between curvature and matter for an homogeneous and isotropic Universe. As an application, we study the case of Abell 586, a relaxed and spherically symmetric galaxy cluster, assuming some matter density profiles.
\end{abstract}

\maketitle

 \section{Introduction}
\label{sec:intro}
It is well established that General Relativity describes several gravitational phenomena at the solar system with very high accuracy, and predicts astrophysical objects like black holes \cite{GR0, GR}. Nevertheless, on galactic and cosmological scales, matching observations requires two unknow components: a non-baryonic form of matter, dark matter, that explains galactic rotation curves and a dynamical mass on galaxy clusters, and an exotic form of energy to explain the late-time accelerated expansion of the Universe, namely dark energy. These two dark components constitute nearly $95 \%$ of the energy content of the Universe and their nature is still a mystery.

On the other hand, several alternative gravitational theories have been proposed to account for the observations usually explained by the presence of dark matter and dark energy, such as, for instance, $f\left(\mathsf{R}\right)$ theories of gravity \cite{fr1, fr2, fr3}.  Another interesting possibility involves the non-minimal coupling between curvature and matter \cite{modelo} (see Refs. \cite{cosm1, cosm2} for proposals in the context of cosmology). The later has a rich lore of theoretical and observational implications, and has bearings on issues such as stellar stability \cite{stellar}, preheating after inflation \cite{inflation}, mimicking of dark matter in galaxies \cite{DM mim1} and clusters \cite{DM mim2} and the large scale effect of dark energy \cite{DE mim1} (see Ref. \cite{review} for a review).

An important tool to study gravitationally bound systems is the virial theorem. In the cosmological context, it is associated with the Lazyer-Irvine equation \cite{Irvine, Layzer, zel}, or the generalised cosmic virial theorem as it is often refered to. This equation can be directly applied to gravitationally collapsed astrophysical objects at different scales. If the gravitationally bound object is sufficiently relaxed, then that equation reduces to the usual virial theorem. From the deviation of the measured quantities such as mass, radius and velocity dispersion relatively to the virial ratio, we are able to test the existence of extra matter or the effect of modified gravity. These relations have been used to study the interaction between the dark components of the Universe in galaxy clusters as Abell 586 and Abell 1689 \cite{abell1,abell1.5, balfagon, abell2}, of dark energy components leading to structure formation \cite{wang, eu}, in the context of $f(R)$ gravity \cite{k} and scalar-tensor theories \cite{winther}, and for modified gravitational potentials of the form $\varphi\left(a, \left|\overrightarrow{r_1}-\overrightarrow{r_2}\right|\right)$, where the cosmological evolution appears in terms of the scale factor,  $a\left(t\right)$ \cite{yuri}.

In this work, we address the problem of adapting the Layzer-Irvine equation to theories with non-minimal coupling between matter and curvature.

The work is organised as follows. First, we shortly review the non-minimally coupled matter-curvature model \cite{modelo} and some of its unique properties. Then, we derive the Layzer-Irvine equation for these theories following up the procedure outlined in Refs. \cite{Layzer, wang, winther}. In section 5, we apply the obtained Layzer-Irvine equation on the Abell 586, a relaxed spherically symmetric galaxy cluster, that has not undergone any relevant merging process in the last few Gyrs \cite{cypriano}. Finally, we show how to estimate the velocity dispersion potential assuming that the cluster is in hydrostic and virial equilibrium, for different matter density profiles.

 \section{Non-minimal curvature-matter coupling}
We start by considering the gravity model with a non-minimal coupling between curvature and matter as expressed by the functional action \cite{modelo}:

\begin{equation}
\label{action}
S=\int\left[\frac{1}{2}f_{1}\left(\mathsf{R}\right)+\left(1+f_{2}\left(\mathsf{R}\right)\right)\mathcal{L}_{m}\right]\sqrt{-g}d^{4}x ,
\end{equation}

\noindent
 where $f_{1}\left(\mathsf{R}\right)$   and $f_{2}\left(\mathsf{R}\right)$   are arbitrary functions of the Ricci scalar, $\mathsf{R}$, and $g$  is the metric determinant. We point out that by setting $f_{1}\left(\mathsf{R}\right)=2\kappa\mathsf{R}$   and $f_{2}\left(\mathsf{R}\right)=0$   we recover General Relativity, where $\kappa=c^4 / 16\pi G$,  $G$ being the Newton's gravitational constant.
Varying the action with respect to the metric yields the field equations ~\cite{modelo}:

\begin{equation}
\label{field equations}
F\mathsf{R}_{\nu}^{\mu}-\frac{1}{2}\delta_{\nu}^{\mu}f_{1}-\left(g^{\mu\sigma}\nabla_{\sigma}\nabla_{\nu}-\delta_{\nu}^{\mu}\square\right)F=\left(1+f_{2}\right)T_{\nu}^{\mu} ,
 \end{equation}
 
\noindent with $F_{i}\equiv df_{i}/d\mathsf{R}\;\left(i=1,2\right)$, $F\equiv F_{1}+2F_{2}\mathcal{L}_m$  , and $T_{\mu \nu}$   is the energy-momentum tensor of matter.

The Biachi identities for the Einstein tensor, $\nabla_{\mu}G_{\nu}^{\mu}=0$  , imply for the above expression:
\begin{equation}
\nabla_{\mu}T_{\nu}^{\mu}=\left(\mathcal{L}_m\delta_{\nu}^{\mu}-T_{\nu}^{\mu}\right)\nabla_{\mu}\ln\left(1+f_{2}\right) .
\end{equation}

This is one of the fundamental features of the model (\ref{action}) - the non-conservation of the energy-momentum tensor. This property induces an extra force acting on a test particle, which is orthogonal to the fluid four-velocity  and can be expressed for a perfect fluid as:

\begin{equation}
f^{\mu}=\frac{1}{\rho + p}\left[\frac{F_2}{1+f_2}(\mathcal{L}_m + p)\nabla_{\nu}\mathsf{R}+\nabla_{\nu}p\right]h^{\mu\nu} ,
 \end{equation}

\noindent where $h^{\mu\nu}=g^{\mu\nu}+u^{\mu}u^{\nu}$ is the projection operator.

\section{Perturbed Fridemann-Lemaître-Robertson-Walker model}

We now proceed in order to derive the Layzer-Irvine equation for the non-minimal coupling model described by the action Eq. (\ref{action}). To do so, we follow closely the derivation performed in Refs. \cite{Layzer, winther, wang, abell1}.

We consider that the Universe is well described by a perfect fluid, $T^{\mu\nu}=\left(\rho+p\right)u^{\mu}u^{\nu}+pg^{\mu\nu}$  , where $u{}^{\mu}=\left(1,u^{i}\right)$   is the four-velocity under the condition $u^{\mu}u_{\mu}=-1$  . We also admit an homogeneous and isotropic spacetime described by the Robertson-Walker metric,$\gamma_{ij}$, whose perturbations are given by the line element

\begin{equation}
ds^{2}=-\left(1+2\Phi\right)dt^{2}+a^{2}\left(t\right)\left(1-2\Psi\right)\gamma_{ij}dx^i dx^j .
\end{equation}

From now on, we consider the choice of the Lagrangian density as $\mathcal{L}_m=-\rho$ (see Ref. \cite{lagrangian choices} for a thorough discussion) Defining the potential velocity in terms of the spatial components of the 4-velocity as $u_{i}=-\partial_{i}v$   and computing the first order perturbation in the components $\delta T_{0}^{i}$   of the stress tensor for a matter dominated epoch, $\rho\approx\rho_{m}$  , and pressureless Universe, we get \cite{frazao}

\begin{equation}
\dot{v}+\dot{\Phi}_{c}v=\Phi+\delta\Phi_{c} ,
\end{equation}

\noindent where $\Phi_{c}=\ln\left(1+f_{2}\right)$  . This expression can be rewritten in terms of the four-velocity as

\begin{equation}
\label{4-velocity}
\dot{u}_{i} = - \nabla_r(\Phi + \delta \Phi_c -v\dot \Phi_c) .
\end{equation}

We shall make the assumption that the flow velocity associated to the expansion rate of the Universe is much smaller than the typical peculiar velocities of cosmic structures. Then $u_{i}\approx a\dot{x}_{i}\equiv v_{m\: i}$  . Under this condition, Eq. (\ref{4-velocity}) can be expressed in a more convenient form

\begin{equation}
\label{velo}
\frac{\partial}{\partial t}\left(av_{m}\right)=-a\nabla_r\left(\Phi+\delta\Phi_{c}-\dot{\Phi}_{c}v\right) .
 \end{equation}

The evolution of matter density perturbations is given in the Fourier space by \cite{frazao}

\begin{equation}
\dot{\delta\rho_{m}}+3H\delta\rho_{m}=3\dot{\Psi}\rho_{m}-\left(\frac{k^{2}}{a^{2}}\frac{v}{a}\right)\rho_{m} ,
 \end{equation}

\noindent where $H=\dot{a} / a$  is the expansion rate. In the real space, using the notation $\sigma_{m}\equiv\delta\rho_{m}$  , only considering peculiar velocities  and in the subhorizon approximation ($k/a>H$) we can write

\begin{equation}
\label{pert evol}
\dot{\sigma}_{m}+3H\sigma_{m}=-\frac{1}{a}\nabla_{x}\cdot\left(\rho_{m}\overrightarrow{v_{m}}\right) .
 \end{equation}

Finally, from the time component of the non-conservation of the energy-momentum tensor, for a pressureless $\left( w=0\right)$ Universe with Lagrangian density $\mathcal L =-\rho_m$, then \footnote{\unexpanded{\makeatletter\let\@bibitemShut\relax\makeatother}%
  The generalisation of the previous result is as follows \cite{modified friedmann}
  \[
    \dot{\rho}_{m}+3H\left(1+w\right)\rho_m=\frac{F_{2}}{1+f_{2}}\left(\alpha-1\right)\rho_{m}\dot{\mathsf{R}} ,
  \]\protect\trick.
where $\alpha=\begin{cases}
1\;\;\:, & \mathcal{L}=-\rho_{m}\\
-w, & \mathcal{L}=p
\end{cases}$  so that the Lagrangian density has the form $\mathcal{L}=-\alpha\rho_{m}$   (see Ref. \cite{lagrangian choices} for a discussion) and $w=p/\rho_{m}$   is the equation of state parameter. 
}

\begin{equation}
\label{friedmann}
\dot{\rho}_{m}+3H\rho_m=0 .
\end{equation}

\section{The Layzer-Irvine equation}
We are now able to derive the Layzer-Irvine equation. We start by contracting Eq. (\ref{velo}) with $a\overrightarrow{v}_{m}\rho_{m}d^{3}r$  , for $r=ax$  , and then integrating over the volume, we get:

\begin{equation}
\label{LI initial}
\int\rho_{m}a\overrightarrow{v}_{m}\frac{\partial}{\partial t}\left(a\overrightarrow{v_m}\right)d^{3}r = -\int a^{2}\overrightarrow{v_m}\rho_{m}\nabla_r\left(\Phi+\delta\Phi_{c}-\dot{\Phi}_{c}v\right)d^{3}r .
 \end{equation}

Using Eq. (\ref{friedmann}), the left hand side of Eq. (\ref{LI initial}) can be expressed as $\frac{\partial}{\partial t}\left(a^{2}K\right)$,  where $K\equiv1/2\int\rho_{m}v_{m}^{2}d^{3}r$   is the kinetic energy associated with the peculiar velocity.

The right hand side can be evaluated through an intergration by parts:

\begin{equation}
\begin{aligned}
&-\int a^{2}\overrightarrow{v_m}\rho_{m}\nabla_r\left(\Phi+\delta\Phi_{c}-\dot{\Phi}_{c}v\right)d^{3}r = \\
&=-\int\nabla_r\left(a^{2}\overrightarrow{v_m}\rho_{m}\left(\Phi+\delta\Phi_{c}-\dot{\Phi}_{c}v\right)d^{3}r\right)+\int\left(\Phi+\delta\Phi_{c}-\dot{\Phi}_{c}v\right)\nabla_{r}\cdot\left(a^{2}\overrightarrow{v_m}\rho_{m}\right)d^{3}r \\
&=-\int\left(\Phi+\delta\Phi_{c}-\dot{\Phi}_{c}v\right)a^{2}\left(\dot{\sigma}_{m}+3H\sigma_{m}\right)d^{3}r .
\end{aligned}
\end{equation}

We have used the fact that the first integral vanishes since it corresponds to a total derivative and we have resorted to Eq. (\ref{pert evol}).

Collecting the results, we get

\begin{equation}
\label{LI almost}
\frac{\partial K}{\partial t}+2HK  =
- \int  (\Phi + \delta \Phi_c -\dot\Phi_cv) \frac{\partial}{\partial t}\left(\sigma_md^3r\right) .
\end{equation}

We will require that each potential satisfies Poisson's equation. We use some of the results of Ref. \cite{Layzer}. First of all, let us define the autocorrelation function $f\left(\overrightarrow{r}\right)$ of the matter density perturbation field, $\sigma_m$, as

\begin{equation}
\left<\sigma_m(\overrightarrow{r},t) \sigma_m(\overrightarrow{r'},t)\right>\:=\:\left<\sigma_m ^2\right>f\left(|\overrightarrow{r}-\overrightarrow{r'}|\right) .
\end{equation}

From which we can define some scales. We should also note that $\left<\sigma_m\left(\overrightarrow{r},t\right)\right>=0$.
Secondly, we use that

\begin{equation}
\label{hubble}
\frac{\partial}{\partial t} \frac{1}{|r-r'|} = - \frac{H}{|r-r'|} .
\end{equation}

Since we require that the potentials to satisfy the Poisson's equation, then any of them can be expressed in terms of the matter density perturbation as

\begin{equation}
\varphi= -G\int \frac{\sigma_m(r',t)}{|r-r'|}d^3r' .
\end{equation}

Bearing this in mind, the right hand side of Eq. (\ref{LI almost}) can be expressed as

\begin{equation}
\label{pot}
\begin{split}
-\int \varphi \frac{\partial}{\partial t} (\sigma_m d^3r) = G\int \frac{\partial}{\partial t} (\sigma_m d^3r) \int \frac{\sigma'_m}{|r-r'|}d^3r'
\\  =G\int \frac{\partial}{\partial t} (\sigma'_m d^3r') \int \frac{\sigma_m}{|r-r'|}d^3r ,
\end{split}
\end{equation}

\noindent where $\sigma_m \equiv \sigma_m(\overrightarrow{r},t)$ and $\sigma'_m \equiv \sigma_m(\overrightarrow{r'},t)$. Now, recalling the result (\ref{hubble}), the expression (\ref{pot}) can be written as

\begin{equation}
G\int \frac{\partial}{\partial t} (\sigma'_m d^3r') \int \frac{\sigma_m}{|r-r'|}d^3r = - (\dot U_{\varphi} + HU_{\varphi}) ,
\end{equation}

\noindent where
\begin{equation}
\label{potential}
U_{\varphi}\equiv -\frac{G}{2}\int \int \frac{\sigma_m \sigma'_m}{|r-r'|}d^3r d^3r' = \frac{1}{2} \int \varphi \:\sigma_m d^3r .
\end{equation}

Note that in the case of clusters, the non-minimal couplings effects on the gravitational coupling \cite{frazao} are negligible, hence $G_{eff} \approx G$. Now we can write the Layzer-Irvine equation in the form

\begin{equation}
\frac{\partial }{\partial t}(K+U_{\Phi} + U_{\delta\Phi_c-\dot \Phi_c v})+H(2K+U_{\Phi} + U_{\delta\Phi_c-\dot \Phi_c v}) =0 ,
\end{equation}

\noindent which can rearranged into a more convenient form

\begin{equation}
\label{virial}
\frac{\partial }{\partial t}(K+U +U_{NMC})+H(2K+U +U_{NMC}) =0 ,
\end{equation}

\noindent with $U\equiv U_{\Phi}$ and 

\begin{equation}
U_{NMC} \equiv U_{\delta\Phi_c-\dot \Phi_c v} =\frac{1}{2} \int \left(\delta\Phi_c-\dot \Phi_c v\right) \: \sigma_m d^3r .
\end{equation}

For a relaxed astrophysical system which no longer evolves in time, we get a generalised virial theorem for the theories with non-minimal coupling between curvature and matter:

\begin{equation}
2K+U+U_{NMC}=0 .
\end{equation}

From this equation we can analise cluster of galaxies and impose some constraints on the non-minimal model. Clearly, any deviation from the usual virial ratio can be expressed as:

\begin{equation}
\frac{U_{NMC}}{U} = -2 \frac{K}{U} - 1 .
\end{equation}

\section{The Abell 586 cluster}
We consider now the well known relaxed cluster Abell 586, following up the procedure developed in Refs. \cite{abell1,abell2}. We assume the obvious cases of the top-hat and isothermal spheres density profiles. In order to test the sensitivity os the results, we adopt tentatively the Navarro-Frenk-White (NFW) density profile \cite{NFW}, even though this is known to be somewhat unaccurate for clusters. As we shall see, results for $U_{NMC}$ are dependent on the density profile choice, even though not strongly so. It is relevant to bear in mind that the considered density is  exclusively baryonic.

\subsection{Top-hat density profile}
In this case, one assumes that the kinetic and potential energy densities are well described by \cite{abell1}

\begin{equation}
\rho_K \simeq \frac{9}{8\pi} \frac{M}{R^3} \sigma_v ^2 ,
\end{equation}

\begin{equation}
\rho_W \simeq -\frac{3}{8\pi} \frac{G\: M^2}{\left<R\right>R^3} ,
\end{equation}

\noindent where $M$ e R are the total baryonic mass and radius of Abell 586 (galaxies and intra-cluster gas), $\sigma_v$ is the velocity dispersion and $\left<R\right>$ is the mean intergalactic radius. The virial ratio is simply given by

\begin{equation}
\frac{K}{U} \equiv \frac{\rho_K}{\rho_W}=- 3\frac{\sigma_v ^2 \left<R\right>}{G\:M} .
\end{equation}

\subsection{Navarro-Frenk-White density profile}
The Navarro-Frenk-White model is very useful in realistic N-body simulations. It is characterised by the energy density \cite{NFW}:

\begin{equation}
\label{nfw}
\rho(r)=\frac{\rho_0}{\frac{r}{r_0}\left(1+\frac{r}{r_0}\right)^2} ,
\end{equation}

\noindent where $r$ is the distance from the centre, $\rho_0$ and $r_0$ are the density and shape parameters, respectively.
The total mass and mean radius can be computed by integrating Eq. (\ref{nfw}) over the volume, as described in Ref. \cite{abell2}:

\begin{equation}
M = 4\pi \int_0 ^R \rho(r)r^2 dr = 4\pi r_0 ^3 \rho_0\left [\ln\left(1+\frac{R}{r_0}\right)-\frac{R}{R+r_0}\right],
\end{equation}

\begin{equation}
\left<R\right> = r_0 \frac{\left[\frac{R}{r_0}-2\ln\left(1+\frac{R}{r_0}\right)+\frac{R}{R+r_0}\right]}{\left[\ln\left(1+\frac{R}{r_0}\right)-\frac{R}{R+r_0}\right]} .
\end{equation}

We point out that $r_0$ can be numerically calculated from the mean radius, $\left<R\right>$. Thus, the density parameter, $\rho_0$, is immediatly solved numerically. From these quantities we can now estimate the kinetic and potential energy densities assuming a constant average velocity distribution \cite{abell2} :

\begin{equation}
\rho_K = \frac{9}{8\pi}\frac{M}{R^3}\sigma_v ^2 ,
\end{equation}

\begin{equation}
\rho_W = - \frac{3GM^2}{4\pi R^3 r_0}     \frac{\left[\left(1+\frac{R}{r_0}\right)\left[\frac{1}{2}\left(1+\frac{R}{r_0}\right)-\ln\left(1+\frac{R}{r_0}\right)\right]-\frac{1}{2}\right]}              {\left[\left(1+\frac{R}{r_0}\right)\ln\left(1+\frac{R}{r_0}\right)-\frac{R}{r_0}\right]^2} .
\end{equation}

Since the case we are studying has spherical symmetry, the total volume is simply $V=4\pi R^3/3$, and the ratio between total peculiar kinetic and potential energies is the same as the ratio of the energy densities:

\begin{equation}
\frac{K}{U} \equiv \frac{\rho_K}{\rho_W} =  -\frac{3}{2} \frac{\sigma_v ^2r_0}{G\:M} \frac{\left[\left(1+\frac{R}{r_0}\right)\ln\left(1+\frac{R}{r_0}\right)-\frac{R}{r_0}\right]^2}       {\left[\left(1+\frac{R}{r_0}\right)\left[\frac{1}{2}\left(1+\frac{R}{r_0}\right)-\ln\left(1+\frac{R}{r_0}\right)\right]-\frac{1}{2}\right]} .
\end{equation}

\subsection{Isothermal density profile}
Another useful density profile is the isothermal density profile, given by

\begin{equation}
\label{isothermal}
\rho \left(r\right)= \frac{\rho_0}{\left(\frac{r}{r_0}\right)^2} .
\end{equation}

Since there is no characteristic scale in this case, we set the fiducial parameters, $r_0$ and $M_0 = 4\pi \rho_0 r_0 ^3 /3$, as the total mass and radius of the halo. Therefore, the mass and the mean radius are \cite{abell2}:

\begin{equation}
M=4\pi \int_0 ^R \frac{\rho_0 r^2}{\left(\frac{r}{r_0}\right)^2}dr=M_0 \frac {R}{r_0} ,
\end{equation}

\begin{equation}
\left<R\right> = \frac {R}{2} .
\end{equation}

With these quantities, we can now get the expressions for the peculiar kinetic and potential energy densities, assuming constant average velocity dispersion \cite{abell2}

\begin{equation}
\rho_K=\frac{9}{8\pi}\frac{M}{R^3}\sigma_v ^2 ,
\end{equation}

\begin{equation}
\rho_W =  -\frac{3GM^2}{4\pi R^4} .
\end{equation}

The virial ratio can be then straightforwardly obtained

\begin{equation}
\frac{K}{U} \equiv \frac{\rho_K}{\rho_W} =- \frac{3}{2} \frac{\sigma_v ^2R}{G\:M} .
\end{equation}

\subsection{Analysis}

\begin{table}
\centering
  \begin{tabular}{| l | r| }
    \hline 
    Method & $\sigma_v (km/s)$ \\ \hline\hline 
    X-ray Luminosity & $1015 \pm 500$  \\ \hline
    X-ray Temperature & $1174 \pm 130$  \\ \hline
    Weak Lensing & $1243 \pm 58$ \\ \hline
    Velocity distribution & $1161 \pm 196$ \\ \hline
  \end{tabular}
\caption{Velocity dispersion data from different observation methods of Abell 586 as given by Ref. \cite{cypriano}.}
\label{table:sigma}
\end{table}

For the analysis of Abell 586 we use data from Ref. \cite{cypriano}, namely:

\begin{itemize}
\item total baryonic mass is given by $M_{bar}=M_{gas}\left(1+0.16\:h_{70}^{0.5}\right)$, where $M_{gas}=0.48\times10^{14}\:M_{\odot}$ is the intercluster gas mass, and $h_{70}=H_0 / 70$ is the reduced Hubble parameter at present;
\item radius, $R=422\:kpc$;
\item velocity dispersion obtained from different methods as shown in Table \ref{table:sigma}.
\end{itemize}

From the 31 galaxies of A586, we can compute the averaged distance from a galaxy $i$ with equatorial coordinates $\left(\alpha_i, \delta_i\right)$ to the centre of the cluster $\left(\alpha_c, \delta_c\right)$ throughout the formula \cite{abell2}

\begin{equation}
r_i ^2 = 2 d^2 [1-cos(\alpha_i -\alpha_c)cos(\delta_c)cos(\delta_i)-sin(\delta_c)sin(\delta_i)] .
\end{equation}

\noindent where $d$ is the radial distance from the centre of the cluster to Earth.

Thus, we get
\begin{equation}
\left<R\right>=223.6\:kpc .
\end{equation}

Furthermore, the errors are computed through the propagation uncertainties formula for $f\left(x_i\right)=\prod_i x_i ^{n_i}$,

\begin{equation}
\Delta f = \left|f \right| \sqrt{\sum_i \left(\frac{n_i\Delta x_i}{x_i}\right)^2} , 
\end{equation}

\noindent where the $\Delta$ symbol denotes the standard deviation for each measurable quantity.

With the density profiles from the previous section, we can now compute the ratio $U_{NMC}/U$ and detect the deviations from the standard virial ratio $K/U=-1/2$.  In Table \ref{table:U_NMC} we exhibit several values for the mentioned ratio according to different density profiles and observational sources. As in Ref. \cite{abell2}, the non-minimal ratio, and the ensued virial ratio, yields higher values for weak lensing velocity dispersion. This observational method is not the most reliable one since it introduces correlation between estimated of mass and velocity, as pointed out in Ref. \cite{abell2}. In our case, we want to identify deviations from the baryonic virial ratio and interpret them as an effect from the non-minimal coupling between curvature and matter.

\begin{table}
\centering
  \begin{tabular}{| l | c | c | c| }
    \hline 
    $U_{NMC}\;/\;U$ & Top-hat & NFW &  Isothermal \\ \hline\hline 
    X-ray Luminosity & $4.8 \pm 5.7$ & $5.3 \pm 6.2$ & $4.5 \pm 5.4$   \\ \hline
    X-ray Temperature & $6.7 \pm 1.8$ & $7.4 \pm 2.0$ & $6.3 \pm 1.7$   \\ \hline
    Weak Lensing & $7.7 \pm 1.1$ & $8.5 \pm 1.2$ & $7.2 \pm 1.0$ \\ \hline
    Velocity distribution & $6.6 \pm 2.6$ & $7.3 \pm 2.9$ &  $6.1 \pm 2.5$\\ \hline
  \end{tabular}
\caption{NMC constraints for different density profiles in terms of various observational methods.}
\label{table:U_NMC}
\end{table}

\indent Keeping in mind Eqs. (\ref{potential}) and (\ref{virial}), we can express the non-minimal potential in terms of the function $f_2(R)$ of Eq. (\ref{action}) as

\begin{equation}
U_{NMC} = \frac{1}{2} \int d^3r \sigma_m \left(\frac{F_2}{1+f_2}\right) \left[ \delta \mathsf{R} \:c^2 - \dot {\mathsf{R}} v\right] .
\end{equation}

The scalar curvature can be computed by performing the trace of Eq. (\ref{field equations}), assuming a general power law coupling function $f_2\left( \mathsf{R} \right)=\left( \mathsf{R}/\mathsf{R}_n \right)^n$, yielding \cite{galaxies, gal}:

\begin{equation}
\mathsf{R}=\frac{1}{2\kappa}\left[1+\left(1-2n\right)\left(\frac{\mathsf{R}}{\mathsf{R}_n}\right)^n \right]\rho-\frac{3n}{\kappa}\square \left[\left(\frac{\mathsf{R}}{\mathsf{R}_n}\right)^n\frac{\rho}{\mathsf{R}}\right] .
\end{equation}

\noindent Considering the weak coupling regime, $(\mathsf{R}/\mathsf{R}_n)^n \ll1$, the above expression simplifies to

\begin{equation}
\mathsf{R}\approx\frac{\rho}{2\kappa} ,
\end{equation}

\noindent which is consistent with the assumption of the subhorizon approximation, where $\mathsf{R}\sim H^2$. The Ricci scalar fluctuation is then $\delta \mathsf{R} \approx \delta \rho/2\kappa$, whilst the time derivative, $\dot{\mathsf{R}}$, can be calculated using Eq. (\ref{friedmann}), yielding $\dot {\mathsf{R}} \approx \dot \rho /2\kappa \approx -3H\rho/2\kappa$.

Hence, under these conditions, the non-minimal coupling can be expressed as
\begin{equation}
U_{NMC} = \frac{1}{2} \int d^3r \sigma_m \frac{n}{\rho} \left(\frac{\rho}{2\kappa \mathsf{R}_n}\right)^n \left[ \sigma_m \:c^2 + 3H\rho v\right] .
\end{equation}

For each different density profile, we have an estimate for the value of the non-minimal coupling potential energy. Additionally, from Ref.  \cite{DM mim2} the best fit value of the index $n$ for Abell 586 is $n=0.43$. Thus, it results that $\mathsf{R}_{0.43}\equiv 1/\sqrt{r_{0.43}}\approx 5.69\times 10^{-8}$ m$^{-1/2}$, since the characteristic lenght for A586 is $r_{0.43}\sim 0.01$ pc \cite{DM mim2}. From here, we are able to estimate the velocity potential of the cluster.

Since the cluster is virialised, we shall assume that each constituent galaxy has the same peculiar velocity, $\left< v_m \right> = v_m$. And from the previous definition and the isotropy and spherical symmetry of the cluster, it follows that

\begin{equation}
v_m = -\partial_r v \implies v = -  \sigma_v r + v_0 ,
\end{equation}

\noindent where $v_0$ is the velocity potential at $r=0$. From this expression, and for each value of velocity dispersion given by various observational methods and for each density profile we obtain a well defined value of $v_0$. The results are shown in Table \ref{table:v_0}. Clearly, the value $v_0$ depends on the density profile one chooses. We note that this quantity is merely an integration constant, which has no particular physical meaning.

\begin{table}
\centering
  \begin{tabular}{| l | c | c | c| }
    \hline 
    $v_0 \left(kpc^2 s^{-1}\right)$ & Top-hat & NFW &  Isothermal  \\ \hline\hline 
    X-ray Luminosity & $-2.377\times 10^{16}$ & $-1.581\times 10^{16}$ & $-1.095\times 10^{15}$   \\ \hline
    X-ray Temperature & $-3.347\times 10^{16}$ & $-2.216\times 10^{16}$ & $-1.548\times 10^{15}$   \\ \hline
    Weak Lensing & $-3.812\times 10^{16}$ & $-2.520\times 10^{16}$ & $-1.765\times 10^{15}$ \\ \hline
    Velocity distribution & $-3.263 \times 10^{16}$ & $-2.160\times 10^{16}$ &  $-1.509\times 10^{15}$\\ \hline
  \end{tabular}
\caption{NMC constraints on the parameter $v_0$ for different density profiles in terms of the various observational methods.}
\label{table:v_0}

\end{table}

\begin{figure}[ht!]
\centering
\includegraphics[width=90mm]{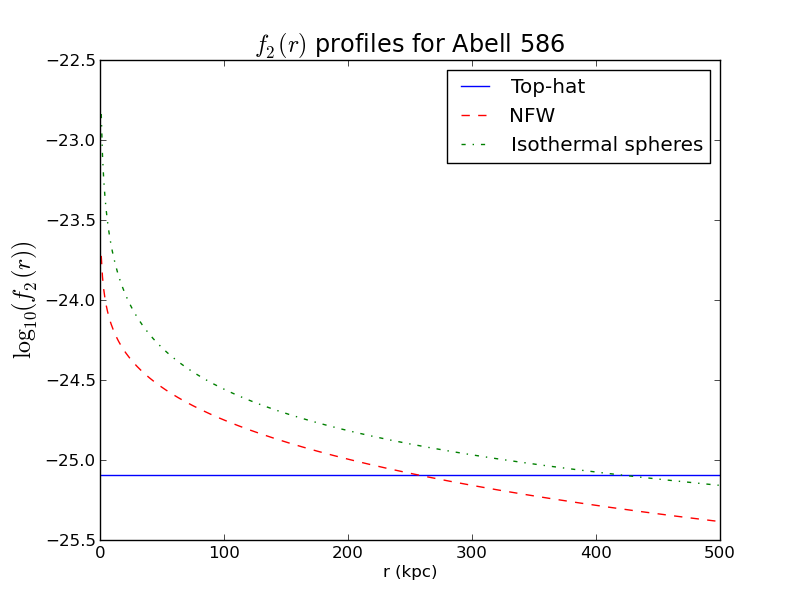}
\caption{Function $f_2\left(r\right)$ for the cluster A586 and in terms of each density profile.}
\label{f2}
\end{figure}

We can also analyse the $f_2\left(\mathsf{R}\right)$  behaviour for A586. This coupling function can be express as:

\begin{equation}
f_2 \left(\mathsf{R}\right) =  f_2\left(r\right)=\left( \frac{\rho\left(r\right)}{2\kappa \mathsf{R}_{0.43}}\right)^{0.43} ,
\end{equation}

\noindent where $\rho\left(r\right)$ is the density profile. In Fig. \ref{f2} we show the plot of the function $f_2$ in terms of the distance $r$ from the cluster's centre.

Notice that the isothermal spheres and the Navarro-Frenk-White density profiles are singular at $r=0$, which results in a much stronger coupling near the centre of A586, whilst the top-hat profile exhibits a constant effect of the matter-curvature coupling all over the cluster's size.

\section{Conclusions} In this work, we have derived the Layzer-Irvine equation in the context of alternative theories of gravity with non-minimal coupling between curvature and matter, as established by the action functional Eq. (\ref{action}). We consider a pressureless matter dominated Universe and the power-law function, $f_2\left(\mathsf{R}\right)=\left(\mathsf{R}/\mathsf{R}_n\right)^n$, with $n=0.43$ \cite{DM mim2}, in the sub-horizon approximation  $k^2 \ll a^2H^2$. We have also chosen for the Lagrangian density $\mathcal{L}=-\rho$.

We find that in these theories, as far as that generalised cosmic virial theorem is concerned, an extra potential energy term appears as a result of the non-minimal coupling.

We have analised the case of the spherically symmetric and relaxed cluster A586, where the virial ratio is computed for only baryonic matter. The role of the extra potential energy is crucial. Indeed, using the velocity dispersion values obtained from various observational methods and three different density profiles (top-hat, Navarro-Frenk-White, and isothermal), we find that the ratio between the non-minimal coupling potential energy and the baryonic energy potential, $U_{NMC}/U$, to be of the order of $\sim7$. We also concluded, as in previous work, Ref. \cite{abell2}, that the velocity dispersion value from X-ray luminosity is not very reliable.

We have then estimated the peculiar velocity potential as a linear function of the distance from the cluster centre, given that the A586 has already reached the virial and hydrostatic equilibrium. We point out that in general we could expect a sum of power-law terms with a velocity profile of the form $v_m = \sum_{\alpha} v_{\alpha} r^{\alpha}$, which leads to a complicated issue of computing the leading coefficients $v_{\alpha}$. Despite of that, future observational improvements might allow for extracting velocity profiles that are consistent with these theories with non-minimal coupling and confront them with the velocity field based on General Relativity and a suitable dark matter distribution. 

Finally, we have analysed the $f_2\left(\mathsf{R}\right)$ function over the distance from the cluster's centre for the different density profiles used in this work, concluding that for singular density profiles at $r=0$, the coupling function is naturally stronger.

\vspace{1,5cm}

{\it \bf Acknowledgements:}
The work of one of us (O.B.) is partially supported by Fundação para a Ciência e Tecnologia (Portugal) under the project PTDC/FIS/111362/2009.


\end{document}